\documentclass{article}

\usepackage{arxiv}

\usepackage[utf8]{inputenc} 
\usepackage[T1]{fontenc}    
\usepackage{hyperref}       
\usepackage{url}            
\usepackage{booktabs}       
\usepackage{amsfonts}       
\usepackage{nicefrac}       
\usepackage{microtype}      
\usepackage{lipsum}
\usepackage{graphicx}
\graphicspath{ {./images/} }

\usepackage[version=3]{mhchem}

\usepackage{multicol}

\title{Self-preserving ice layers on \ce{CO2} clathrate particles: implications for Enceladus, Pluto and similar ocean worlds}

\author{Mathias Bostr\"om \\
 Centre for Materials Science and Nanotechnology\\
 Department of Physics, University of Oslo\\ 
 NO-0316 Oslo, Norway\\
  \And
 Victoria Esteso \\
Multifunctional Optical Materials Group\\
Instituto de Ciencia de Materiales de Sevilla  \\ 
Calle Am\'erico Vespucio 49, 41092 Sevilla, Spain\\
\And
 Johannes Fiedler\thanks{Department of Physics and Technology, University of Bergen, All\'egaten 55, 5007 Bergen, Norway} \\
 Physikalisches Institut\\
 Albert-Ludwigs-Universit{\"a}t Freiburg\\
 Hermann-Herder-Str. 3, 79104 Freiburg, Germany\\
   \And
    Iver Brevik \\
 Department of Energy and Process Engineering\\
 Norwegian University of Science and Technology\\
 NO-7491 Trondheim, Norway\\
  \And
   Stefan Yoshi Buhmann\\
  Institut f\"ur Physik, Universit\"at Kassel\\
  Heinrich-Plett-Str. 40, 34132 Kassel, Germany\\
  \And
   Clas Persson \\
  Centre for Materials Science and Nanotechnology\\
 Department of Physics\\
 University of Oslo\\ 
 NO-0316 Oslo, Norway\\
    \And
    Sol Carretero-Palacios \\
Departamento de F\'isica de Materiales\\
Universidad Aut\'onoma de Madrid\\
28049 Madrid, Spain\\
    \And
 Drew F. Parsons \\
 Department of Chemical and Geological Sciences\\
University of Cagliari\\
Cittadella Universitaria, 09042 Monserrato, CA, Italy \\
  \And
R. W. Corkery\thanks{Department of Applied Mathematics, Research School of Physics, ANU College of Science, Oliphant Building 60, Mills Road, The Australian National University, Canberra Acton ACT 2601, Australia }\\
Surface and Corrosion Science\\ 
Department of Chemistry\\ 
KTH Royal Institute of Technology\\
SE 100 44 Stockholm, Sweden\\ 
 \texttt{corkery@kth.se}\\
}

\begin{document}
\maketitle
\begin{abstract}
 Under both engineering and natural conditions on Earth and in the Universe, some gas hydrates (e.g., \ce{CO2} clathrates) are found to be 
stabilised outside their window of thermodynamic stability by 
formation of an ice layer---a phenomenon termed self-preservation. Low 
occupancy surface regions on type I \ce{CO2} clathrate structures 
together with the self-preserving ice layer lead to an effective 
buoyancy for these structures which restricts the size range of type I 
\ce{CO2} clathrates particles that float in the ocean on
Enceladus, Pluto and similar oceanic worlds.
Our goal here is to investigate the implications of Lifshitz 
forces and low occupancy surface regions on type I \ce{CO2} clathrate 
structures for their self-preservation through ice layer formation, presenting a plausible model based on  multi-layer interactions through 
dispersion forces. 
We use optical data and theoretical models for the dielectric 
response for water, ice and gas hydrates with different occupancy. 
Taking this together with the thermodynamic Lifshitz free energy, we 
model the energy minima essential for formation of ice layers at the 
interface between gas hydrate and liquid water.
We predict that the growth of an ice layer between 0.01 and 
0.2\,$\mu$m thick on \ce{CO2} clathrate surfaces depends on the presence of surface regions in the gas hydrates with low occupancy of gas molecules. In addition, the effective particle density is estimated delimiting a range of particle size/compositions that would be buoyant in different oceans. 
Over geological time, deposition of floating \ce{CO2} hydrate particles could result in the accumulation of kilometre-thick gas hydrate layers above liquid water reservoirs, and below the water ice crusts of their respective ocean worlds. On Enceladus, destabilisation of near-surface \ce{CO2} hydrate deposits could lead to increased gas pressures that both drive plumes and entrain stabilised hydrate particles to be be re-deposited on the surface of Enceladus or ejected into the E-ring of Saturn. Furthermore, on ocean worlds such as Enceladus and particularly Pluto, the accumulation of thick \ce{CO2} hydrate deposits could insulate its ocean against freezing. In preventing freezing of liquid water reservoirs in ocean worlds, the presence of \ce{CO2} hydrate layers could enhance the habitability of ocean worlds in our solar systems and on the exoplanets and exomoons beyond. 
\end{abstract}

\section{Introduction}

Gas hydrates are important materials on Earth as well as on  many planets and moons in our solar system~\cite{MethaneHydratesinsolarsystem}. The ice-capped ocean-bearing moons Enceladus and Europa are two of the best examples beyond Earth where gas hydrates can be formed in their respective salty oceans~\cite{EnceladusBook}. Liquid water, gas hydrates, and complex molecules are potentially favourable for life within the Solar System and on exoplanets~\cite{Nimmo2016,LeviExoplanets,EnceladusBook,KhawajaPostbergHillierKlennerKempfNolleReviolZouSrama2019,Harsono_2020_AandA}. Another example is Mars, where methane co-exists with subterranean water implying the existence of methane hydrates~\cite{Fonti2010}.  

Gas hydrates (also known as clathrates) are crystalline structures comprising water molecules arranged in cages that enclathrate guest gas molecules within~\cite{doi:10.1021/acs.energyfuels.0c01291}. As discussed by~\cite{TakeyaRipmeester2008}, these caged guest molecules are essential for formation for hydrates, and the structure is thermodynamically stable only above a certain occupancy of gas molecules. These structures are furthermore stable only within a limited range of temperatures and pressures. As one example, \ce{CH4} gas hydrates turn out to be stable below the freezing temperature of water (e.g., at atmospheric pressure and temperatures between 240\,K to 273\,K)~\cite{TakeyaRipmeester2008}. The solution to this unexpectedly wide stability range seems to be that a thin, typically 1-2\,$\mu$m~\cite{Falenty2009}, water ice layer forms on the surface~\cite{TakeyaRipmeester2008,Falenty2009}. Some gas hydrates (\ce{CH4}, \ce{CO2} and \ce{N2} type I hydrates, but not \ce{H2S} hydrates~\cite{TakeyaRipmeester2008}) are notably seen to be supported outside their normal window of thermodynamic stability by formation of an ice layer---this phenomenon is termed ``self-preservation'' or ``anomalous preservation''~\cite{TakeyaRipmeester2008,Falenty2009}, and has been observed in shallow permafrost~\cite{Shakhova2017,Chuvilin,Shakhova2019}.
 
The formation of an ice layer on the hydrate is not only driven by kinetics as suggested by~\cite{Falenty2009}, but can also be driven by Lifshitz-force mediated equilibrium thermodynamics~\cite{Bostrom_acsearthspacechem.9b00019}. Indeed, if the self-preserving ice layer effect on hydrates is spontaneous, as we find, this adds further weight to the argument that gas hydrates may be stabilised in a way that was not previously understood.

The aim of this work is to present a model that offers a potential explanation to the occurrence of micron-sized water ice layers on \ce{CO2} gas hydrate surfaces in cold water. Fundamentally, such layers of water ice on gas hydrates, if present, are important for exploration and production of fossil fuels and are vital for understanding the potential for methane contributions to greenhouse gases as the planet becomes warmer. We also point to the relevance of self-preservation ice layers in pipes with flowing natural gases, in seabed~\cite{doi:10.1002/2016GL068656}, sediments~\cite{doi:10.1029/2007GC001920}, and in permafrost~\cite{NaturalGas2003}. In the current context, the role of Lifshitz energy can be important: the induced ice films on water may change the buoyancy of micron-sized \ce{CO2} gas hydrate particles.

   \begin{figure}
   \centering
   \includegraphics[width=0.4\columnwidth]{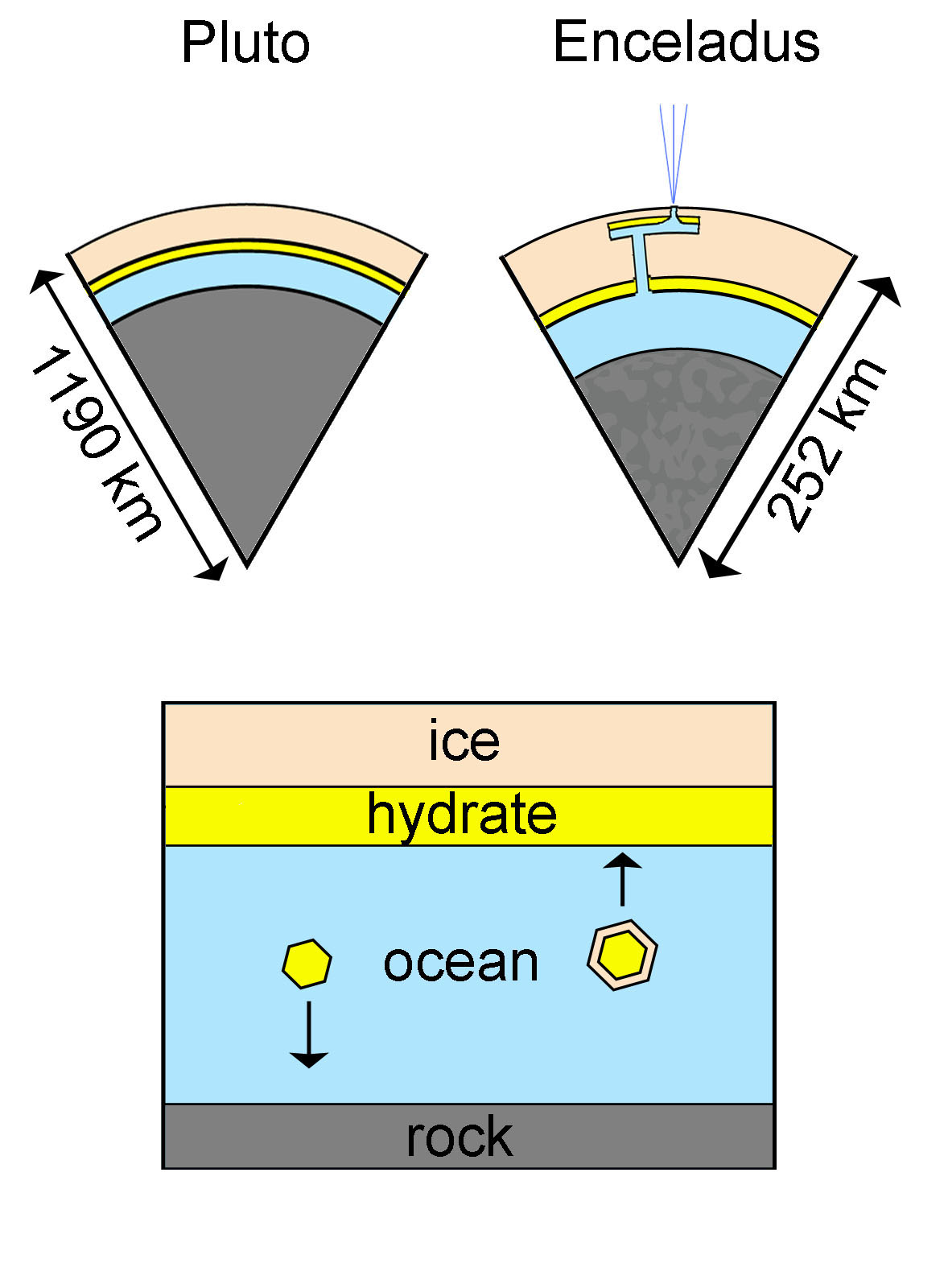}
      \caption{Figure shows schematically the relative size and internal structure of Pluto and Enceladus, both with proposed rocky cores, liquid water oceans, capping hydrate and ice layer. For Enceladus, we also illustrate the subsurface shallow reservoirs of liquid water proposed by~\cite{Matson2018}. For Pluto, we illustrate the structure with the thermally-insulating layer of \ce{CO2} hydrate, similar to the methane hydrate layer proposed by~\cite{KamataNimmoSekineKuramotoNoguchiKimuraTani2019}. As illustrated, in the bottom part of the figure, ice stabilised \ce{CO2} hydrate particles can float in the ocean layer towards the hydrate layer, whereas hydrates without an ice stabilising coat sink towards the rock layer. In both the case of Enceladus and Pluto, we propose that significantly thick \ce{CO2} hydrate deposits may form at the upper ocean surface, impacting the structure, heat and mass transport evolution of the body.}
         \label{fig:SchematicEnceladus}
   \end{figure}

We consider ice formation on \ce{CO2} clathrates and its potential impact in different Solar System ocean worlds. Presence of both liquid water and gaseous \ce{CO2} on the icy moon Enceladus have been inferred from measurements by the Cassini spacecraft~\cite{EnceladusBook}. Other ocean world candidates in the Solar System include Callisto, Europa,  Ganymede and Titan~\cite{Nimmo2016}. Even the Kuiper Belt Objects (KBOs) Pluto and Triton have been proposed to have a reservoir of liquid water~\cite{Sternaad1815,Nimmo2016}. These icy Solar System bodies have been in some focus  for exploration and search for potential candidates to host extraterrestrial life~\cite{Nimmo2016} in liquid water oceans.  It is not unrealistic to assume a large number of habitable ocean worlds even when only a small fraction of the Universe is considered. Here we are interested in two such potential ocean worlds: Enceladus and Pluto.
Depending on the material properties and sizes, ice-coated hydrate particles may enter the fissures in the Enceladean ice crust rather than sinking to the bottom of the ocean. From there hydrates may enter fracture systems that feed Enceladus' complex plume without destabilising. If they do not become trapped in deposits within the icy crust, hydrate particles may be directly ejected from the surface to the E-ring or to fall back onto the surface of the south polar region. Notably, floating particles have sizes of the same order of magnitude as the observed ice particles in the E-ring around Saturn~\cite{EnceladusBook}. These could come from composite water ice particles but also from less abundant self-preserving clathrate hydrates decomposed by solar winds into ice particles and gas molecules. Stabilised hydrates trapped in potentially large amounts under or within the icy crust during transport may form deposits that may reactivate under changed conditions and have a delayed participation in the plume activity. 
 As schematically indicated in Fig.~\ref{fig:SchematicEnceladus} our model predicts that nanoscopic to microscopic ice coatings are essential for the \ce{CO2} hydrate particles to be able to float in oceans. We propose that these floating particles could in some ocean worlds  lead to thick \ce{CO2} hydrate deposits forming below or in the ice crust, thereby redistributing \ce{CO2} upwards, impacting the structural, thermal and dynamical properties of the bodies. We propose that \ce{CO2} deposits above the Enceladean ocean, particularly those closer to the surface, may be more prone to destabilisation and associated pressure driven mass ejection via plumes and seeps.  We also propose that the insulating hydrate layers on Pluto proposed and modelled by~\cite{KamataNimmoSekineKuramotoNoguchiKimuraTani2019} will likely not be comprised of pure methane hydrates as suggested, but a mixture of ice-stabilised \ce{CO2} hydrates with additional components involving \ce{CH4} hydrates, mixed \ce{CO2}-\ce{CH4} hydrates, and other hydrates. We expect the general ideas presented here to be relevant for countless exoplanets beyond our Solar System~\cite{LeviExoplanets,StuartRossTaylor1999}.

 \section{Computational Details}
 
Following an idea presented by~\cite{Elbaum}, according to which  intermolecular dispersion forces can be a vital contribution to ice premelting, ~\cite{Bostrom_acsearthspacechem.9b00019} recently investigated how these forces can also cause ice-cold water to freeze at some interfaces. Our present work considers a system with gas hydrate--ice--water, and we will replace an old model for the dielectric functions for ice-cold water with an improved new one. In order to model the energies involved in ice formation at \ce{CO2} clathrate-water interfaces we require the dielectric functions for ice, water, and \ce{CO2} clathrate structures.

 The first material we need to model is the \ce{CO2} gas hydrates. At specific temperatures and pressures, known as quadruple points, clathrate structures can be formed when gas molecules are present in ice-cold water~\cite{doi:10.1029/94GL01858}. Modelling of dielectric functions for \ce{CO2}  clathrates were presented in a previous work~\cite{Bostrom_acsearthspacechem.9b00019}. In this letter, the material modelling is expanded to the full range of different gas molecular occupancy in the clathrate structure. This turns out to be an essential step and we predict here that a (surface) region with low occupancy of gas molecules is essential to obtain agreement with the experimentally observed ice layers discussed above~\cite{TakeyaRipmeester2008,Falenty2009}.  
 
Secondly, the dielectric model for ice proposed by~\cite{Elbaum} remains  reliable in the frequency range required for calculations of dispersion forces~\cite{LUENGO2019}. Thirdly, two new parameterisations~\cite{JohannesWater2019,LUENGO2019} for optical properties for water give similar Hamaker constants for ice-water related systems. We show in this work that by replacing the classic model for the dielectric function of ice-cold water~\cite{Elbaum} with the new model~\cite{JohannesWater2019} for cold water near the quadruple point of \ce{CO2} clathrates, we obtain substantially better agreement with experiments. The difference between the \cite{Elbaum} model for water and the new water models is that the new one relies on optical data from a much wider frequency range than before, and also benefits from improved parameterisation techniques. Especially, the model for cold water presented by~\cite{Elbaum} has been seen to underestimate the strength of the dielectric function of cold water in optical/UV frequencies, primarily due to unavailability of data at X-ray frequencies above 20 eV. The new model from~\cite{JohannesWater2019} includes all these new data. 

 Any realistic theory for self-preservation of gas hydrate must as a critical test be able to accommodate water ice layers of this kind.  Within the old model used by~\cite{Bostrom_acsearthspacechem.9b00019}, only nanosized water ice layers were found on fully occupied \ce{CO2} particles. No ice layers were predicted with lower occupancy of the gas molecules. However, our main result when using the improved water model is that we find a much more realistic micron-sized layer, on low-occupancy, to close-to-half filled, gas hydrates. Such low occupancy regions could occur near the surface due to diffusion of gas molecules. The stability of the system is then controlled by the slow diffusion rate for gas molecules through the ice layer.  


\begin{figure}[t]
   \begin{minipage}[c]{.49\textwidth} 
      \includegraphics[width=\columnwidth]{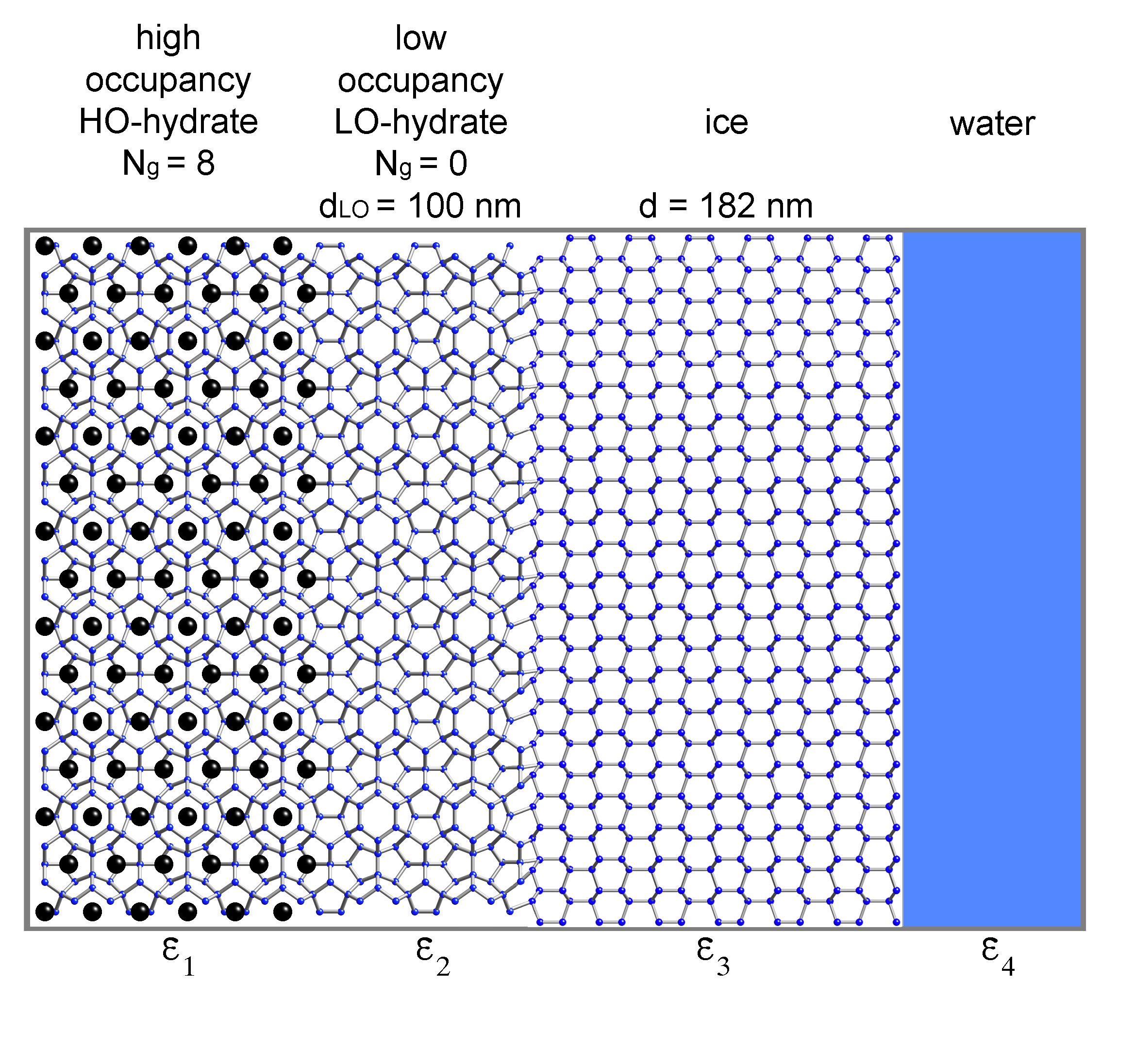}
      \caption{ Schematic diagram of the 4-layer \ce{CO2} hydrate system: high-occupancy (HO) gas  hydrate $\varepsilon_1$, low-occupancy (LO) hydrate $\varepsilon_2$, pure \ce{H2O} ice, $\varepsilon_3$, and pure liquid \ce{H2O}, $\varepsilon_4$.
            }
         \label{fig:SchematicPlanarLayerModelsystem}
   \end{minipage}
   \hfill
   \begin{minipage}[c]{.49\textwidth} 
   \centering
      \includegraphics[width=0.8\columnwidth]{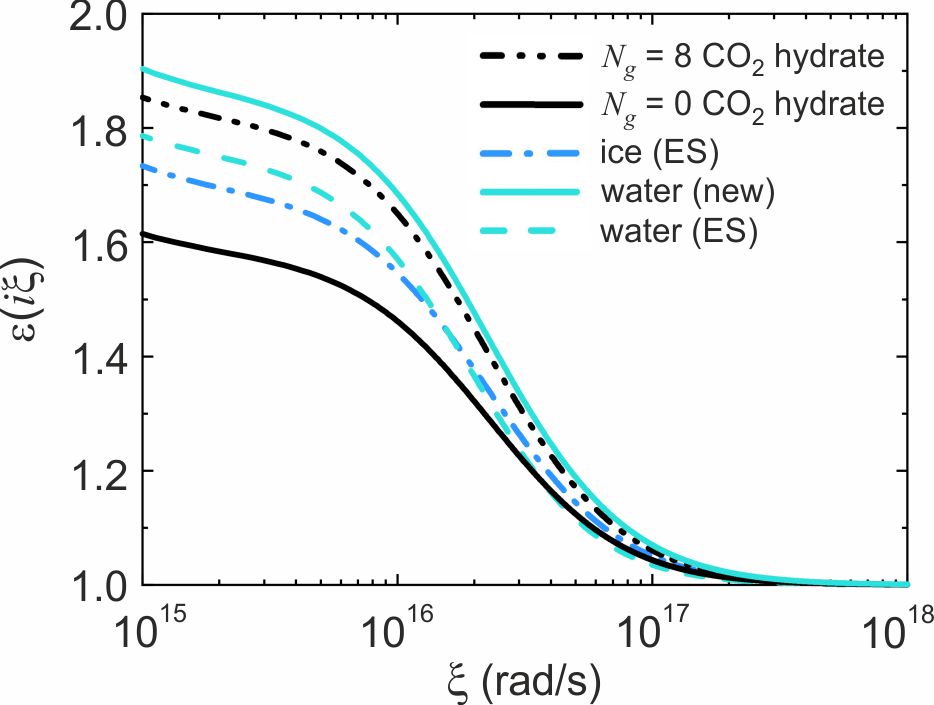}
      \caption{The dielectric functions at {T=273.1\,K} evaluated at imaginary frequencies for different materials: fully occupied ($N_{ \rm g}$ = 8) \ce{CO2} gas hydrate (black dash dot dot line), empty ($N_{ \rm g}$ = 0) \ce{CO2} gas hydrate structure (black solid line), ice-cold water~\cite{JohannesWater2019} (light blue solid line), and ice~\cite{Elbaum} (blue dash dot line). Here $N_{\rm g}$ is the number (between 0 and 8) of gas molecules per 46 water molecules, in the gas hydrate structures.
      The static values $\varepsilon(0)$ are 91.5 for ice\,\cite{Elbaum}, 87.8 in the model for water~\cite{JohannesWater2019}, 16.6 for empty \ce{CO2} structure ($N_{ \rm g}$ = 0), and 25.4 for fully occupied   \ce{CO2} structure ($N_{ \rm g}$ = 8).  For comparison, we also show the older dielectric function for ice-cold water\,\cite{Elbaum} (light blue dash line).
              }
         \label{fig:dielectricfunctions}
   \end{minipage}
\end{figure}
 
The equilibrium ice layer comes from the free energy minimum for the layered structure shown in the schematic Fig.~\ref{fig:SchematicPlanarLayerModelsystem}.   Lattice mismatches often occur at the interfaces between different gas hydrates, between same gas hydrates with different densities, and between water ice and the gas hydrate. Such heterogeneous structures are known to exist in nature, and in fact ice layers on gas hydrates are observed but not fully understood~\cite{TakeyaRipmeester2008,Falenty2009,Nguyen2015}. \cite{Nguyen2015} presented a path to understand this problem using molecular dynamics simulations for a gas hydrate--ice--air system and found that such transition layers are disordered, and can typically be few (commonly 2--3) water layers thick. The disordered interfacial transition layer between ice and gas hydrates when present relieves the pressure of mismatch and aids in understanding nucleation of gas hydrates on ice in the presence of gas. The effective thicknesses for such transitional layers are often negligible and hence needs not to be explicitly modelled. Hence the transitional layer does not affect our results, other than noting that its existence is critical in any model aiming to explain ice formation on gas hydrates be it either in gas or water.


We present in Fig.~\ref{fig:dielectricfunctions} the new dielectric function for ice-cold water~\cite{JohannesWater2019}, together with the model for ice from Elbaum and Schick~\cite{Elbaum,Ref12}. We also show the dielectric function for \ce{CO2} gas hydrate particles with empty (low occupancy) and fully (high occupancy) occupied structures. In our calculations for ice formation at the interface of a gas hydrate in contact with ice-cold water, we use a specific quadruple point, which for \ce{CO2} hydrate is at $T = 273.1\, \rm{K}$ and $p = 12.56 \,\rm{bar}$~\cite{Sloan}. Away from the quadruple point, additional thermodynamical effects must be considered. The model for the dielectric functions for gas hydrates was outlined by~\cite{Bostrom_acsearthspacechem.9b00019}. Ultimately, the equilibrium ice layer thicknesses on \ce{CO2} clathrates will be shown to depend on the gas occupancy, where $N_{\rm g}$ (between 0 and 8) is the number of gas molecules per 46 water molecules, in the gas hydrate structure ($N_{\rm g}$\,=\,8 corresponds to fully occupied). The requirement for stable ice layers to form is that the Lifshitz energy for a layered system with gas hydrate--ice--water has short range repulsion that switches over to long range attraction with an energy minimum at a specific ice layer thickness.
In general formation of both gas hydrate structures and self-preservation ice layers depend on various additional factors, including the amount of dissolved gases~\cite{NINHAM201725,doi:10.1021/acs.langmuir.0c00117}.

For premelting of ice surfaces~\cite{Wettlaufer}, but not ice formation electrostatic energies are important. While electrostatic energies are large near ice--water interfaces they are independent of the ice layer thickness except for atomically thin ice layers and do not contribute to the interaction energy considered in this work~\cite{Bostrom_acsearthspacechem.9b00019}. One pathway reported by~\cite{TakeyaRipmeester2008} to synthesise ice coated self-preserved hydrate particles, involves partial degassing of the outer part of a pure gas hydrate in an inert gas atmosphere or in vacuum during warming. Subsequent conversion of the outer emptied clathrate cells can be made to cubic, then hexagonal, ice and eventually annealed to form a continuous ice gas diffusion barrier on the underlying hydrate particle. In this case the source of the water for the ice layer originates from the particle. Formation of self preserved hydrates in salt water was reported by~\cite{Prasad}. They confirmed self-preservation up to the melting point of ice in the three component system water--gas--NaCl for salt concentrations below a threshold of 0.5-1.5\% NaCl. Self preservation was inhibited at higher salt concentrations, including that of typical present day sea water. 

Salt effects in ice systems were recently studied by~\cite{Thiyam2018} to model how ice particles below a certain size can be trapped under a water surface. It is well known that ice melting, e.g. in Arctic regions~\cite{Armitage2020}, can lead to substantial local reduction of the salt content in sea water. We argue that it is likely that, for example, the water below the ice cap on the moon Enceladus, as well as in the geyser channels that are leading the stream of water, gas, and particles out across the same ice cap, could have a much lower salt concentration compared to the deeper sea water. 
The limiting salt concentration for ice formation is linked to the lowering of the melting temperature of water ice in contact with salt water~\cite{Hall1988NaClEffects}. Formation of ice layers in salt water is expected below a salt specific temperature.  Some additional effects of salt ions on ice formation will be discussed near the end of this work.

 \section{Dispersion Free Energies Across the Ice Layer}

One kind of energy  always contributing is the dispersion energy, modelled for planar multilayered systems by Lifshitz and co-workers~\cite{Dzya}. This Lifshitz energy has found an increasing importance in technological applications such as microelectromechanical and nanoelectromechanical components (MEMS and NEMS), as these components gradually become smaller~\cite{DelRio2005,Stange2019}. Dispersion forces are also used  by living organisms, such as geckos  walking in the ceiling~\cite{Autumn2000}. The Lifshitz dispersion interaction  has important implications for astrophysics.  Here we suggest that the formation of ice layers induced by dispersion energies influences the influx and composition of icy particles via plumes onto  the frozen surface of Enceladus and into the E-ring of Saturn. Furthermore, we propose that large amount of floating ice-coated \ce{CO2} hydrate particles could potentially result in an insulating gas hydrate layer above liquid water reservoirs at the lower edge of the icy surface on Enceladus and other icy Solar System bodies. Activity involving liquid water, with a proposed role for insulating methane/carbon monoxide hydrate layers~\cite{KamataNimmoSekineKuramotoNoguchiKimuraTani2019}, has been inferred as a possibility to explain varying crater abundances (i.e.\ different surface ageing) on the icy dwarf planet Pluto~\cite{Sternaad1815,Nimmo2016}. One of the reasons why the  \ce{CO2} molecule was discarded as a source for a potential insulating layer on Pluto \cite{KamataNimmoSekineKuramotoNoguchiKimuraTani2019} was that its gas hydrate structure has a density which is too high to float, and rather than accumulate at the lower edge of the surface ice cap, the gas hydrates would sink to the bottom.  As we point out in our result section, this could change in the presence of self preservation ice layers.

We consider gas hydrates covered with an ice layer that are sufficiently large to be treated as locally planar. In this model, the surface force acting on the ice--water interface determines how the ice layer grows or melts. The corresponding potential is given by the Lifshitz free energy, whose density per unit area at temperature $T$ reads as~\cite{Ellingsen_2007,Esteso4layerPCCP2020}
\begin{equation}
F(d) =\frac{k_{\rm B}T}{2 \pi}{\sum_{m=0}^\infty}^\prime \int\limits_0^\infty \mathrm d k^\parallel \, k^\parallel \sum_{\sigma=\rm {TE, TM}}  \ln[1-\tilde r_{\sigma}^{32}r_{\sigma}^{34}
 \mathrm e^{-2\kappa_3^\perp d}]\,, \label{100}
\end{equation}
where ${k}^\parallel$ is the component of the wave vector ${\bf k}=(2\pi/\lambda)\hat{\bf k}$ parallel to the surface, and $\kappa_i^\perp= \sqrt{{k^\parallel}^2+\varepsilon_i\xi_m^2/c^2}$ (with $i=1,\dots,4$ numerating the layer according to Fig.~\ref{fig:SchematicPlanarLayerModelsystem}, $\xi_m=m\, 2\pi k_{\rm B}T/\hbar$ the Matsubara frequencies, the reduced Planck constant $\hbar$, the Boltzmann constant $k_{\rm B}$ and the speed of light $c$) is a real quantity as the imaginary part of the wave vector. The primed sum in the expression above denotes that the first term ($m$ = 0) has to be weighted by $1/2$. The photon polarisations are $\sigma=$TE (transverse electric), and $\sigma=$TM (transverse magnetic). The multiple reflections between the interfaces lead to~\cite{Buhmann12a,Ellingsen_2007}, 
\begin{equation}
 \tilde r_\sigma^{32} =\frac{r_\sigma^{32}+r_\sigma^{21}\mathrm e^{-2\kappa^\perp_2 d_{\rm LO}}}{1+r_\sigma^{32}r_\sigma^{21}\mathrm e^{-2\kappa_2^\perp d_{\rm LO}}}\,. \label{reflec}
\end{equation}
One then needs the expressions for the Fresnel reflection coefficients, $r_\sigma^{ij}$, by which a photon in medium $i$ is assumed to be reflected from medium $j$
\begin{equation}
    r_{\rm TE}^{ij} = \frac{\kappa^\perp_i-\kappa^\perp_j}{\kappa^\perp_i+\kappa^\perp_j}\,,  \qquad r_{\rm TM}^{ij} = \frac{\varepsilon_j\kappa^\perp_i-\varepsilon_i \kappa^\perp_j}{\varepsilon_j \kappa^\perp_i+\varepsilon_i \kappa^\perp_j} \,. \label{eq:rtTETM}
\end{equation}
The free energy~(\ref{100}) describes the action on the interface between the low occupancy hydrate layer ($\varepsilon_2$) with thickness $d_{\rm LO}$ and the ice layer ($\varepsilon_3$) with thickness $d$, see Fig.~\ref{fig:SchematicPlanarLayerModelsystem}.

\begin{figure}[t]
   \begin{minipage}[t]{.49\textwidth} 
      \includegraphics[width=\columnwidth]{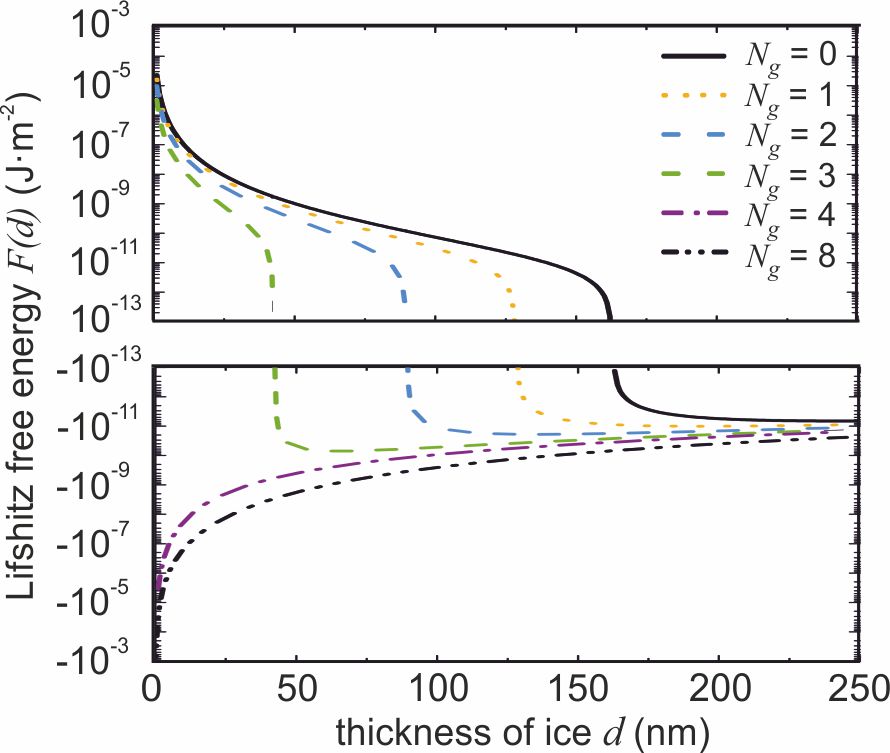}
      \caption{The Lifshitz free energy for water--ice--\ce{CO2} gas hydrates as a function of ice layer thickness using the dielectric function for water from~\cite{JohannesWater2019} and dielectric function for ice from~\cite{Elbaum}.}
         \label{fig:co2hydrate_FDFP}
   \end{minipage}
   \hfill
   \begin{minipage}[t]{.49\textwidth} 
      \includegraphics[width=\columnwidth]{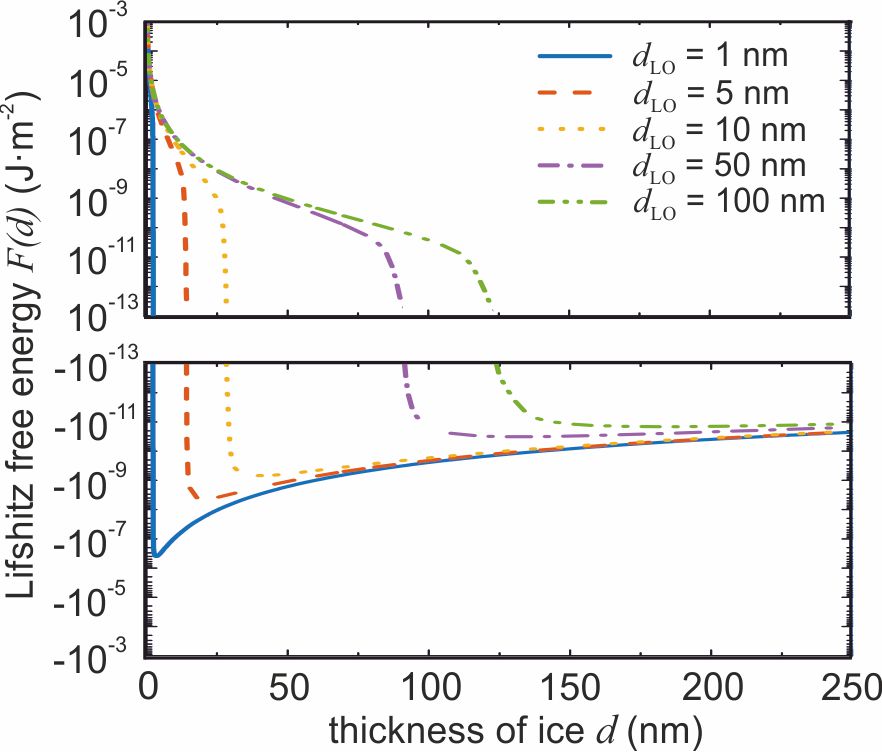}
      \caption{The Lifshitz free energy for water--ice--\ce{CO2} ($N_{ \rm g}$\,=\,0) gas hydrate--\ce{CO2} ($N_{ \rm g}$\,=\,8) gas hydrates as a function of ice layer thickness for different  thicknesses ($d_{\rm LO}$) of the low occupancy shell  over the high occupancy core, using the dielectric function for water from~\cite{JohannesWater2019} and dielectric function for ice from~\cite{Elbaum}.}
         \label{fig:diff_dLO}
   \end{minipage}
\end{figure}

\section{Results: Ice Layers and Buoyancy}

We begin with the earlier unresolved problem~\cite{Bostrom_acsearthspacechem.9b00019}, namely that for experimental conditions the ice film thickness needs to be much thicker than the nanosized water ice films that was predicted in the past~\cite{Bostrom_acsearthspacechem.9b00019}. This result was in contradiction to the well known experimental result discussed in the literature~\cite{TakeyaRipmeester2008,Falenty2009}. Theoretical calculations~\cite{Bostrom_acsearthspacechem.9b00019} applied the average experimental occupancy (not fully 100\% occupied positions for the gas molecules) together with an older model of the dielectric function of ice-cold water~\cite{Elbaum}. This led to the problem that ice films were predicted ranging only up to a few nanometres thick for \ce{CO2} hydrate structures with $N_{\rm g}$ values between 5 and 8. For low occupancy \ce{CO2} hydrate structures no ice layer was predicted. As noted elsewhere, up to micron-sized self-preservation ice layers have been observed experimentally~\cite{TakeyaRipmeester2008,Falenty2009}.



The new case considered here depends on a new improved model for water dielectric function~\cite{JohannesWater2019}. This leads to entirely different predictions for the surface properties of gas hydrates required for self-preserving ice-layers. In Fig.~\ref{fig:co2hydrate_FDFP} we observe that \ce{CO2} clathrates with as low as 0-38\% occupancy (the upper limit corresponds to $N_{\rm g}$\,=\,3) can have ice layers between 65 nm and 245 nm,  corresponding to minima in the Lifshitz free energy. Hence almost micron-sized ice layers can form on the clathrate structures under the strict condition that a sufficient fraction of the gas molecules, at least in the interface region, has escaped into the surrounding liquid water.

Equilibrium ice layer thicknesses on homogeneous \ce{CO2} gas hydrate particles at various gas densities  are predicted to be 245, 193, 136, and 65 nm  for $N_{\rm g}$\,=\,0, 1, 2, and 3, respectively. There is no ice layer for higher occupancy. A key point to observe is that use of  the Elbaum and Schick model for cold water leads to the prediction that only partially full to fully occupied gas hydrate structures at the quadruple point can have thin ($< 4$ nm) ice layers. In sharp contrast, the new water model requires low occupancy ($<38\%$) gas hydrate structures in order to have Lifshitz interaction promoting ice formation. Here, a thick ice layer as thick as 245 nm is predicted. This influences self-preserving layers, the presence/absence of ice on \ce{CO2} gas hydrates, and buoyancy of these gas hydrate particles in cold water.

Of course, empty clathrate structures are not stable, at least not in their bulk forms. A system of great interest is therefore a bulk material (with high occupancy of gas molecules) with a finite surface region where depletion of gas molecule has resulted in a low occupancy region.  Here, the effect of such layered media influences the ice formation in ice-cold water. As an example we consider first a surface region with empty clathrate structure and a bulk region with fully occupied structure. For thin ice layers the free energy shown in Fig.~\ref{fig:diff_dLO}  is not influenced by the bulk region while for very thick ice layers the inner region dominates. The basic theory has been well described in the general multilayered case in the book by~\cite{Pars}. When the low occupancy surface region grows thicker the result eventually approaches that found for homogeneous low occupancy clathrate structures. In Table~\ref{Table_ice_thicknesses}, the ice thickness stabilising the \ce{CO2} gas hydrate is given for a surface region $d_{ \rm LO}$ at two fixed gas occupancy values. We take as examples  a low occupancy surface region $N_{\rm g}$ = 0 and a partially occupied surface region $N_{\rm g}$ = 1. In both cases, the thicker the surface region, the thicker the self-preserving ice layer. One should notice that already very thin surface depleted regions enable the formation of self-preservation ice layers.

    \begin{table}
    \centering

     \begin{tabular}{l|l|l|}
      & \multicolumn{2}{ c| }{Ice thickness, $d$ (nm)} \\    $d_{\rm LO}$ (nm) & $N_{\rm g} = 0$ & $N_{\rm g} = 1$  \\
            \hline
        1 &    3           & 3       \\
        5 &    20          & 16       \\
        10 &   40          & 32       \\
        50 &   131         & 108       \\
        100 &  182         & 149 
    \end{tabular}
      \caption{Self-preserving ice layer thickness $d$ for increasing values of low occupancy (LO) surface region thickness $d_{\rm LO}$ with $N_{ \rm g}$ = 0 (second column). As comparison we also give one example of partially occupied surface region with $N_{ \rm g}$ = 1 (third column). The bulk region is in these examples taken as a fully occupied clathrate structure.}

    \label{Table_ice_thicknesses}
    \end{table}


As a direct application, this could have an important influence on the gas cycle on the moon Enceladus. On this moon, giant plumes of erupted gases via active geysers have been observed and the composition has been measured by the Cassini space craft to be  dominated by water, various salts, and a large number of volatile gases including \ce{CO2}, \ce{CO}, \ce{N2}, \ce{H2S} and \ce{CH4}~\cite{Waite2006,Postberg2011,EnceladusBook}. It has been postulated that  gas hydrates could be important for the creation of the plumes as well as for the chemical composition of the ice layers beneath and in the interior. The plumes are sprayed out from the outer surface of the moon~\cite{Bouquet2015,Matson2012,Matson2018}. \cite{Matson2018} suggested that plumes are driven by \ce{CO2} collecting in a subsurface water table below cap ice less than 1\,km under the surface, and proposed that bubbles and gas dissolved in ocean water are the source of the \ce{CO2}.

Larger molecules (such as \ce{CO2} and \ce{CH4}) form type I clathrate structures, while type II clathrate structures typically host smaller molecules (e.g. \ce{O2} and \ce{N2})~\cite{Sloan}.
 A fully occupied type I \ce{CO2} hydrates has an density greater than water, such that it will sink in water~\cite{McKay2003,PrietoBallesteros2005,Mousis2013,Safi2017}. A layer of water ice can change the effective density and hence the buoyancy of gas hydrate particles~\cite{Bostrom_acsearthspacechem.9b00019,Thiyam2018}. It was recently shown that geometry effects are small for  concentric sphere ice systems when the size of the core is thick enough relative to the thickness of a coating layer~\cite{Prachi_concentricice2019}. This encouraged~\cite{Bostrom_acsearthspacechem.9b00019} to take $d$ as the approximate thickness of each ice film at planar water--\ce{CO2} gas hydrate interfaces with different gas molecular occupancy. Such a theory is expanded below to account for different occupancy (and mass density) in bulk and surface regions for the gas hydrate particle. 

The average density ($\rho_{\rm av}$) of an ice coated gas hydrate particle (approximated as a sphere with a low occupancy surface region) is
\begin{eqnarray}
\rho_{\rm h}&=&{\frac{\rho_{\rm HO} {r_c^3}+\rho_{\rm LO}  \left[r_{\rm h}^3-r_c^3 \right]}{r_{\rm h}^3}}\,, 
\label{Equrhoh}\\
\rho_{\rm av}&=&{\frac{\rho_{\rm h} r_{\rm h}^3+\rho_{\rm i}  \left[(r_{\rm h}+d)^3-r_{\rm h}^3 \right]}{(r_{\rm h}+d)^3}}\,,
\label{Equrho}
\end{eqnarray}
where $\rho_{\rm h}$ is the average mass density of the layered gas hydrate comprising a high occupancy (HO) core with radius $r_c=r_{\rm h}-d_{\rm LO}$ and a low occupancy (LO) surface region with average thickness $d_{\rm LO}$. Similarly, the average mass density of mixed particle comprising an average of the gas hydrate interior ($\rho_{\rm h}$) and a shell of water ice. The mass density of water in pure ice is $\rho_{\rm i}$\,=\,0.9167\,g/cm$^{3}$,~\cite{CRChandbook} giving the number density of water molecules in pure ice as $n_{\rm i}= 3.06\times 10^{-2}$ \,{\AA}$^{-3}$. The gas hydrate corresponds to water molecules and gas molecules in a specific structure with densities given in Table~\ref{RobTable}.

\begin{table}[t]
\centering

 \begin{tabular}{l|l|l}
 $N_{\rm g}$&$n_{\rm M}$ ($10^{-4} $\AA$^{-3}$) &$\rho_O \,(\rm{g/cm}^{3}$) \\
   \hline
0 (empty=LO)&    0           & 0.80 \\
1 &     $5.78$        &0.84         \\
2 &  $11.55$      & 0.88              \\
3 &  $17.33$       & 0.92            \\
4 &  $23.10$     &  0.96           \\
5 &  $28.88$       &  1.01             \\
6 &  $34.65$      &   1.05       \\
7 &  $40.43$      & 1.09        \\
8  (full=HO)  & $46.20$    & 1.13
  \end{tabular}
  \caption{The \ce{CO2} gas hydrates have number densities for water ($n_{\rm wh}$) and gas ($n_{\rm M}$) molecules  at $T=273.1 \,\rm{K}$ with water density in the structure $n_{\rm wh}=2.65\times10^{-2}$ \AA$^{-3}$. The values given are for $N_{\rm g}$ which is the number of gas molecules per 46 water molecules in the gas hydrate structure. We provide the different gas hydrate mass densities ($\rho_O$) for evaluation of effective densities and buoyancies in water. For instance empty, low occupancy (LO), gas hydrate has a density of 0.80\,g/cm$^{3}$. HO = high occupancy (fully occupied).}
\label{RobTable}
\end{table}

\begin{figure}[t]
   \begin{minipage}[c]{.49\textwidth} 
      \includegraphics[width=\columnwidth]{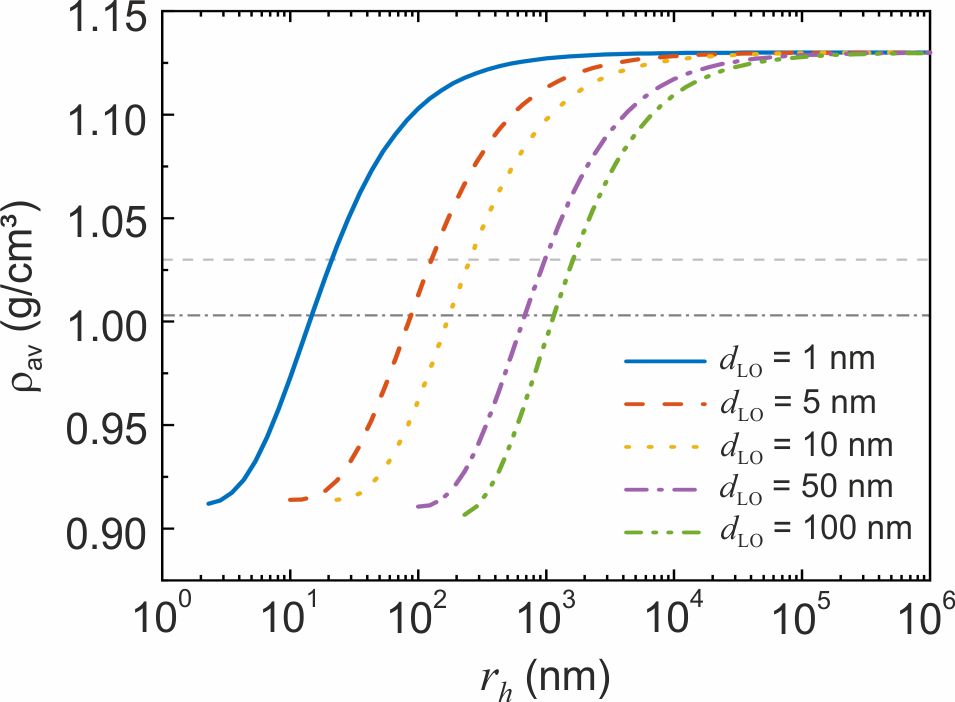}
      \caption{Effective density of particles as a function of gas hydrate radius $r_h$ for different thicknesses $d_{ \rm LO}$ of the outer low-occupancy layer. The corresponding ice thickness for each case is reported in Table~\ref{Table_ice_thicknesses}. For  comparison we also include as dashed and dash-dotted lines assumed densities for the ocean water on Enceladus: ~\cite{Safi2017} estimated it around 1.003 g/cm$^3$, while~\cite{Bouquet2015} estimated it around 1.03 g/cm$^3$. The crossings between the two water density lines and the curves with particle effective  densities are our order of magnitude conclusions for  buoyancy driven particle size float-to-sink transitions.}
         \label{fig:effective_density}
   \end{minipage}
   \hfill
   \begin{minipage}[c]{.49\textwidth} 
   \centering
      \includegraphics[width=\columnwidth]{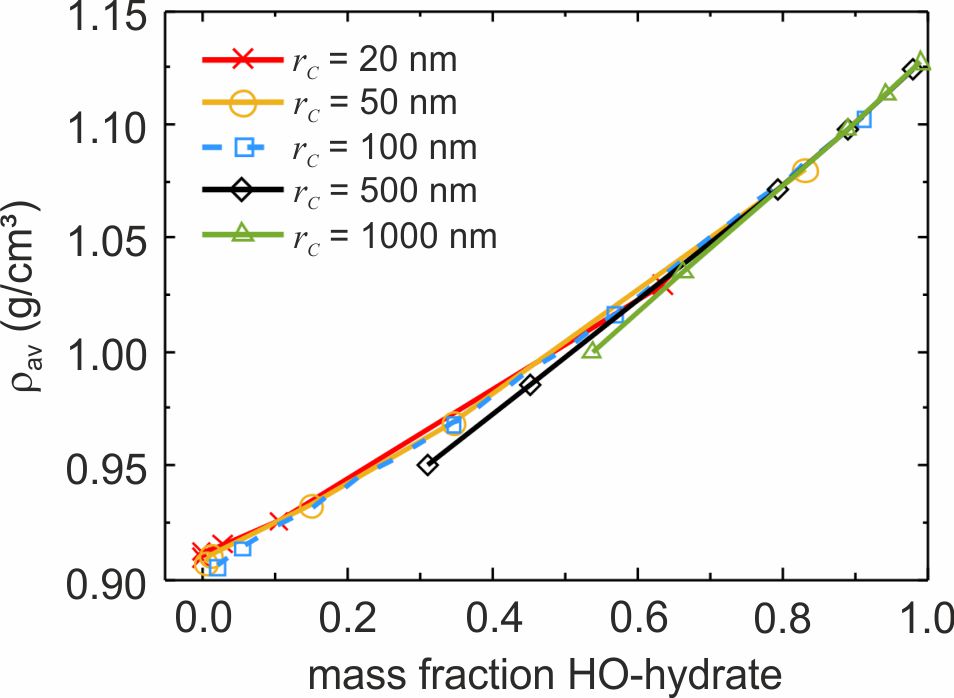}
      \caption{Graph of particle density as a function of high-occupancy-hydrate content in each particle. The five lines correspond to five different high-occupancy-core radii, $r_c=r_{\rm h}-d_{\rm LO}$ = 20, 50, 100, 500 and 1000 nm, with each point in each respective line representing the five scenarios of differing low-occupancy-hydrate thickness given in columns 1 and 2 of Table~\ref{Table_ice_thicknesses}. As can be read directly from the graph, the approximate critical high-occupancy-hydrate mass fraction in particles correspond, in every case, to a critical floating/sinking value of 0.5, below which particles float. We use this as the upper estimate for the high-occupancy-hydrate composition of particles that are able to float in an ocean of density $\rho= 1.0 \,\rm{g/cm}^3$.}
         \label{fig:densty-versus-HO}
   \end{minipage}
\end{figure}


We exploit in Fig.~\ref{fig:effective_density} the equilibrium distances for Lifshitz free energies evaluated above for each combination of surface region thickness and ice layer thickness in the specific example of a fully occupied \ce{CO2} clathrate structure coated with an empty ($N_{ \rm g}$ = 0) clathrate structure. The {0.01-0.2\,$\mu$m} thick ice-layer on homogeneous type I \ce{CO2} hydrate particles that we found above using the new water model suggests that micron-sized instead of nanometre-sized~\cite{Bostrom_acsearthspacechem.9b00019} particles  may float in the ocean under the ice cap on Enceladus. However, both the density of the ice free gas hydrate particle and the additional equilibrium thickness $d$ of the ice layer depend on the internal structure of the gas hydrate particle. Depending on the surface structure of the clathrate particle, the transition from sinking to floating driven by buoyancy  is predicted to occur for typical particle sizes ranging from {0.02} to {2 $\mu$m}.  The critical composition of such particles corresponds to mass fraction of high-occupancy hydrate (in the core of the particle) of around 50\% (see Fig.~\ref{fig:densty-versus-HO}). This number can be used to provide estimates for how the thickness of an insulating hydrate layer is related to the available amount of \ce{CO2} in each specific ocean world. \cite{KamataNimmoSekineKuramotoNoguchiKimuraTani2019} proposed that floating gas hydrates could prevent Pluto's ocean from freezing over geological time by forming a thermally insulating, non-convective layer atop the ocean. They  hypothesised that \ce{CO2} hydrates would sink on Pluto, not accounting for a stabilising ice layer, and therefore they invoked methane hydrates as the insulating hydrate, being naturally buoyant.  We estimate the same layer thickness (10\,km on Pluto) as proposed by~\cite{KamataNimmoSekineKuramotoNoguchiKimuraTani2019} but built from  pure \ce{CO2} hydrates  stabilised by an ice layer with a composition having an upper limit of 50\% pure \ce{CO2} hydrate. This requires around 1\% of the estimated total amount of \ce{CO2} available on Pluto, estimated from \ce{H2O}:\ce{CO2} ratios in the solar Nebula~\cite{Mousis_2009,Eistrupetal_AandA2019} to be bound within the layer. Such a layer on Pluto would be around 50\% less insulating than a pure hydrate due to the mixing with water ice. It would also likely be substantially less viscous than a pure hydrate layer given that the water ice would be the continuous phase, at least initially. Nonetheless, taking into account stabilisation of mixed \ce{CO2}-methane floating hydrates may have important implications for the long term stability of liquid layer on ocean worlds in our Solar System and beyond.

Next, we will give an estimate of the upper limit of thickness of a corresponding hydrate layer on Enceladus. If we consider that all of the \ce{CO2} emitted from the plumes of Enceladus were sourced from a \ce{CO2} hydrate layer on Enceladus, then the integrated thickness over time can be calculated to 1-25 km. In other words, 1-25 km of \ce{CO2} hydrate would have decomposed and been emitted from the surface via plumes. This is based on a rate of plume emission of 100-350\,kg/s~\cite{Postberg2018_book} and a mass fraction of \ce{CO2} in the plume of 0.1-0.8\%~\cite{Waite155} calculated over 4.6 billion years.  If we instead calculate if all \ce{CO2} formed a global hydrate layer,  the thickness of \ce{CO2} hydrate layer before emission would range from 80\,km (taking the \ce{CO2}/water ratio to be cometary~\cite{LeRoyetal_AandA_2015,Eistrupetal_AandA2019}) to 160\,km (taking the \ce{CO2}/water ratio of the to be that of the solar nebula~\cite{Mousis_2009,Eistrupetal_AandA2019}. Clearly given these assumptions, only a small fraction of the total available \ce{CO2} would be required to form significantly thick deposits on Encleladus, or other ocean worlds with similar mass ratios of water:\ce{CO2} and where other sinks of \ce{CO2}, such as carbonation at the ocean/rock interface occur. 
 These are obviously upper limit estimates for the thickness of a gas hydrate layer on Enceladus and does not account for an insulating layer in flux, constantly growing and decreasing by adsorption of gas hydrates and emission of gas hydrates and gas molecules. The important point is that a reasonably thick layer of stabilised hydrate may have formed during Enceladus' lifetime. Such a hydrate layer would have implications for the thermal insulation on the Enceladean ocean via impacting conductive and convective heat transport.

Empty hydrates are generally considered unstable, although these have been isolated under particular laboratory conditions~\cite{JacobsonHujoMolinero_JPCB2009,Falenty2014}, while partially empty hydrates are well known. At the molecular scale, the low-occupancy hydrate layer is therefore considered more likely to be thinner, with structural stabilisation coming from hydrogen bonding with the inner high-occupancy particle core and the outer ice coating. This would tend to favour large populations of buoyant, ice-coated nanometric hydrate particles at the lower end of the distribution, i.e., 20 nm, whose particle number would be 10$^6$ times higher compared to 2 micron particles at the higher end of the distribution, assuming the same number of nuclei. 
Further, at a critical density for buoyancy of 1.003-1.03 g/cm$^{3}$ depending on the ocean density, the 20 nm fraction will have a specific surface area of around 150 m$^2$/g compared with just 1.5 m$^2$/g for 2 micron particles, and at just 0.1\% volume fraction in the sea water, the particle number will be $3\times 10^{19}$ m$^{-3}$. Such particles could enter the plume feed via a sea-spray mechanism~\cite{Postberg2018}. In doing so, these would present a large specific surface area that could act as nucleation sites for ice and salt crystallisation from the liquid phase of the spray droplets and be trapped in the cores of particles ejected to the E-ring or deposited back on the surface of Enceladus~\cite{Postberg2018}.

\section{Discussion}

Our model for anomalous stabilisation of \ce{CO2} gas hydrates via Lifshitz free energy thermodynamical minima may not be directly transferable to the situation for hydrates in the ocean of Enceladus  due to differences in temperature influenced by salinity and under varying pressures at different depths~\cite{Enceladus2018_Ch3_GleinPostbergVance}. Notably, one important effect of salt is to change the freezing point depression of water~\cite{Hall1988NaClEffects,Enceladus2018_Ch3_GleinPostbergVance}.  We stress that the freezing point for water decreases weakly with added salt. While temperatures are very low at the surface of Enceladus the presence of liquid water under the ice cap, notably at an estimated 0.05-0.2 M NaCl concentration and basic pH~\cite{POSTBERG2008438,Postberg2009,Postberg2011,Postberg2018,Glein2020}, indicates ocean water with a temperature close to 273 K~\cite{Enceladus2018_Ch3_GleinPostbergVance}. The estimated water temperature in the ocean on Enceladus is in fact remarkably close to the quadruple point temperature that we use in the present work.~\cite{POSTBERG2008438,Postberg2009,Postberg2011,Postberg2018} (see also~\cite{EnceladusBook}) used the results from a cosmic dust analyser (CDA) to account for three distinct types of ice particles in the plume. The first with almost pure water ice, the second also carried organic compounds, and the third rich in salt. There were also nanoparticles rich in silicon supposed to originate as inclusions in the second and third ice type. One of the most reliable measurements reported by~\cite{Waite155} suggests a true composition (in contrast to  impact-modified composition) with abundant presence of \ce{H2O}, \ce{H2}, and \ce{CO2}. There has also been reported \ce{CH4}, \ce{NH3} and other gases present in measurements~\cite{Waite155}. Comparison between the results from CDA and laboratory experiments further enabled~\cite{POSTBERG2008438,Postberg2009,Postberg2011,Postberg2018} to put limits to the concentrations of different salt species in the ocean on Enceladus.

The freezing point depression is weak at salt concentration levels predicted for the ice-capped ocean on Enceladus. This is important since it suggests that our theoretical model for self-preservation can be relevant also in the presence of salt but at slightly lower temperatures. The added salt ions have at these concentrations only small effects on the dielectric function of water so to an approximation the ice layer thickness would not be dramatically affected by changes in optical properties of the materials. This is also the case for oceans on Earth. Similarly, effects from different NaCl salt concentrations on the melting temperature of \ce{CO2} clathrates were explored by~\cite{Darling1991_NaClEffects}. Improvements of our theory to temperatures away from the quadruple point of the gas hydrate structure can be considered by expanding the recent work from~\cite{Esteso4layerPCCP2020} (see also references therein), while relevant salt effects on ice-water interfaces were discussed by~\cite{Thiyam2018}. Even so, trends are here estimated to be similar in the ocean on Enceladus especially in subsurface water reservoirs just below the water ice cap (where melted water from the ice cap potentially could reduce the salinity). In addition,  low salt conditions are expected  to occur when the water dominated liquid plume mixture passes through holes in the water ice cap coating the ocean. The surface of Enceladus is almost entirely made of pure water ice except near its south pole, in the region of the geysers, especially the region called the {\textquotedblleft}tiger stripes{\textquotedblright}, where there are some light organics, \ce{CO2}, and amorphous and crystalline water ice~\cite{Brown1425}. It seems very likely~\cite{EnceladusBook} that the origin of the \ce{CO2} abundance in this area is linked to the geysers and therefore directly related to the ocean composition.

It is known that the observed blue colour of the Saturn E-ring, in contrast to the red colour of Jupiter's dusty rings or Saturn's G-ring~\cite{Nicholson509,dePater}, suggests water ice particles in the ring having average size of the order of 1$\mu$m~\cite{BAUM198184,Hillier2007,EnceladusBook,Catling}. Measurements by the Cassini spacecraft, within the Cassini-Huygens mission~\cite{EnceladusBook,Hillier2007}, confirmed the size range for particles. They also showed that \ce{CO2}, next after water, is one of the most common species in the plume near the south pole of Enceladus~\cite{EnceladusBook}. \ce{CO2} gas hydrate particles  of different sizes ranging from nanosized to micron-sized can be  ejected~\cite{Matson2018} together with much more abundant water ice particles via plumes onto the surface and partially also into outer space providing material for the E-ring around Saturn.

Our model suggests an alternative mechanism for the source of the \ce{CO2} collecting in the subsurface lakes postulated by~\cite{Matson2018}. They considered bubbles and \ce{CO2} gas dissolved in water as the source of the \ce{CO2} pocket driving the plumes. Our model suggests that buoyant, ice stabilised \ce{CO2} hydrates may  be able to reach the water reservoir below the cap ice and above the water table. Some fraction of these hydrates could then destabilise, for instance through collisions breaking the thinner, stabilising ice layers, or by subsequent infiltration of warmer water  from  hot  spots~\cite{Matson2007,Hsu2015,Bouquet2015}. Such destabilised hydrates could feed either the \ce{CO2} seeps further from the plumes forming \ce{CO2} frost patches, or drive plume activity through pressurisation under the ice cap, resulting in \ce{CO2} and/or \ce{CO2} hydrate particles being ballistically vented to fall back to the surface or to escape into the E-ring. The amount of \ce{CO2} gas in bubbles or dissolved in water is relatively low in contrast to the relatively high amount in \ce{CO2} gas hydrates~\cite{Safi2017}, thus stabilised hydrates could act as a richer source of \ce{CO2} than dissolved gases in the up-welling ocean water.

When the pressures and temperatures of the local environment on Enceladus change at different depths in the ocean, one would expect that it could lead to leakage of gas molecules via diffusion in the interfacial region of each particle. Particles with sizes up to around a few micrometers with a bulk region built by high occupancy type I \ce{CO2} gas hydrates and a substantial interfacial region with low occupancy can have a thin film of water-ice. For such systems, arguments can be made that large particles dominated by bulk regions may sink, while other particles, where the surface region  is thick enough to generate a self-preserving water--ice layer, have the potential to float. With positive buoyancy, these particles could, over geologically long time periods, lead to substantial \ce{CO2} hydrate deposits at the top of the ocean or in the overlying ice crust. From our model, we consequently predict that the fraction of particles that sink through the ocean gradually diminishes through a feedback mechanism. As \ce{CO2} is removed from the ocean by deposition as stabilised hydrates above the ocean, the sinking fraction decomposes near the hot rocky core and, if it does not react with the rocky core material, can be fed back to the ocean to form a new population comprising both sinking and floating hydrates. This cycling could continue until the sinking population becomes depleted through the source being removed to the deposits above the ocean or within the icy crust. 

We have discussed that near-surface deposits in the crust could lead to \ce{CO2} and or hydrate emissions to the surface through seeps or to the surface or E-ring via plumes. As for \ce{CO2} deposits at the top of the ocean, these may accumulate to thick sequences over geological time measured in billions of years. Such layers have been postulated as effective insulators on Pluto as a way to explain why Pluto's ocean has not frozen~\cite{KamataNimmoSekineKuramotoNoguchiKimuraTani2019}. Buoyant hydrates including ice stabilised \ce{CO2} hydrates discussed here may form thick sequences of thermally insulating, high viscosity strata that can limit heat flow to the surface and may be a general phenomena in the hydrosphere of ocean worlds through the proposed stabilisation mechanism.

By floating to the top of the ocean surface, ice-coated \ce{CO2} hydrates can form deposits underneath or within the icy crusts of ocean worlds. Reasonably thick deposits of stabilised hydrates may have formed during the lifetime of Enceladus and Pluto, and in the case of the less tidally disrupted Pluto, these may have formed a continuous global insulator, albeit less thermally insulating than a layer made of pure hydrate. On Enceladus these deposits would be expected to be subject to tectonic and thermal destabilisation. Nonetheless taking into account stabilisation of mixed \ce{CO2}-methane floating hydrates may have important implications for the long term evolution and dynamics of ocean worlds in our Solar System and beyond

  \section{Conclusions}

   \begin{enumerate}
      \item Using the most reliable, experimentally derived dielectric properties of ice and water, we obtain  theoretical results that predict relatively thick, molecular scale ice coats on partially degassed hydrates in ice-cold water pointing to a previously unconsidered mechanism for gas hydrate self-preservation and buoyancy. 
      \item We predict that nano- to micron-sized ice coated type I \ce{CO2} clathrates can float in the water columns of ocean worlds, forming regional or global deposits on top of these oceans, underneath or within the ice layers. 
      \item  \ce{CO2} rich hydrate layers may grow to significant thickness over millions to billions of years. When such layers are global they could provide thermal insulation against freezing of that respective world's ocean, or if local or regional could destabilise to drive and feed ejecta plumes as seen on Enceladus. 
       \item The proposed model presented here for floating \ce{CO2} hydrates is expected to be applicable to the liquid water columns of other ocean worlds within our Solar System, from Europa to Triton and to a significant fraction of the rapidly increasing list of extra solar planetary bodies and their yet to be discovered moons.
   \end{enumerate}

 
    D. F. Parsons acknowledges the grant of resources from the National computational Infrastructure (NCI), which is supported by the Australian Government. We gratefully acknowledge support from the German Research council (grant BU 1803/6-1, S.Y.B. and J.F., BU 1803/3-1, S.Y.B.).
    Last, we would especially like to thank Juan Luengo M\'arquez and Dr. Luis G. MacDowell who informed us that they have independently prepared parameterisations for dielectric functions~\cite{LUENGO2019} that agree to a good approximation with both the new model for cold water~\cite{JohannesWater2019} and the model for ice from~\cite{Elbaum}.

\bibliographystyle{unsrt}  

\bibliography{references}

\begin{thebibliography}{10}

\bibitem{MethaneHydratesinsolarsystem}
Olivier Mousis, Eric Chassefiere, Nils~G. Holm, Alexis Bouquet, Jack~Hunter
  Waite, Wolf~Dietrich Geppert, Sylvain Picaud, Yuri Aikawa, Mohamad Ali-Dib,
  Jean-Luc Charlou, and Philippe Rousselot.
\newblock Methane clathrates in the solar system.
\newblock {\em Astrobiology}, 15(4):308--326, 2015.

\bibitem{EnceladusBook}
P.~M. Schenk, R.~N. Clark, C.~J.~A. Howett, A.~J. Verbiscer, and J.~H. Waite,
  editors.
\newblock {\em {Enceladus and the Icy Moons of Saturn}}.
\newblock The University of Arizona Press, Tucson, 2018.

\bibitem{Nimmo2016}
F.~Nimmo and R.~T. Pappalardo.
\newblock {Ocean worlds in the outer solar system}.
\newblock {\em Journal of Geophysical Research: Planets}, 121:1378--1399, 2016.

\bibitem{LeviExoplanets}
A.~Levi, D.~Sasselov, and M.~Podolak.
\newblock {The Abundance of Atmospheric CO2 in Ocean Exoplanets: a Novel CO2
  Deposition Mechanism}.
\newblock {\em The Astrophysical Journal}, 838(24):1--45, 2017.

\bibitem{KhawajaPostbergHillierKlennerKempfNolleReviolZouSrama2019}
N~Khawaja, F~Postberg, J~Hillier, F~Klenner, S~Kempf, L~Nölle, R~Reviol,
  Z~Zou, and R~Srama.
\newblock {Low-mass nitrogen-, oxygen-bearing, and aromatic compounds in
  Enceladean ice grains}.
\newblock {\em Monthly Notices of the Royal Astronomical Society},
  489(4):5231--5243, 10 2019.

\bibitem{Harsono_2020_AandA}
D.~Harsono, M.~V. Persson, A.~Ramos, N.~M. Murillo, L.~T. Maud, M.~R.
  Hogerheijde, A.~D. Bosman, L.~E. Kristensen, J.~K. J{\o}rgensen, E.~A.
  Bergin, R.~Visser, J.~C. Mottram, and E.~F. van Dishoeck.
\newblock Missing water in class i protostellar disks.
\newblock {\em Astronomy and Astrophysics}, 636:A26, 2020.

\bibitem{Fonti2010}
S.~Fonti and G.~A. Marzo.
\newblock {Mapping the methane on Mars}.
\newblock {\em Astronomy and Astrophysics}, 512(A51):1--6, 2010.

\bibitem{doi:10.1021/acs.energyfuels.0c01291}
Ngoc~N. Nguyen, Mirza Galib, and Anh~V. Nguyen.
\newblock Critical review on gas hydrate formation at solid surfaces and in
  confined spaces—why and how does interfacial regime matter?
\newblock {\em Energy \& Fuels}, 34(6):6751--6760, 2020.

\bibitem{TakeyaRipmeester2008}
Satoshi Takeya and John~A. Ripmeester.
\newblock Dissociation behavior of clathrate hydrates to ice and dependence on
  guest molecules.
\newblock {\em Angewandte Chemie International Edition}, 47(7):1276--1279,
  2008.

\bibitem{Falenty2009}
Andrzej Falenty and Werner~F. Kuhs.
\newblock “self-preservation” of co2 gas hydrates—surface microstructure
  and ice perfection.
\newblock {\em The Journal of Physical Chemistry B}, 113(49):15975--15988,
  2009.

\bibitem{Shakhova2017}
Natalia Shakhova, Igor Semiletov, Orjan Gustafsson, Valentin Sergienko, Leopold
  Lobkovsky, Oleg Dudarev, Vladimir Tumskoy, Michael Grigoriev, Alexey Mazurov,
  Anatoly Salyuk, Roman Ananiev, Andrey Koshurnikov, Denis Kosmach, Alexander
  Charkin, Nicolay Dmitrevsky, Victor Karnaukh, Alexey Gunar, Alexander
  Meluzov, and Denis Chernykh.
\newblock Current rates and mechanisms of subsea permafrost degradation in the
  east siberian arctic shelf.
\newblock {\em Nature Communications}, 8:15872, 2017.

\bibitem{Chuvilin}
E.~Chuvilin, B.~Bukhanov, D.~Davletshina, S.~Grebenkin, and V.~Istomin.
\newblock Dissociation and self-preservation of gas hydrates in permafrost.
\newblock {\em Geosciences}, 8:431, 2018.

\bibitem{Shakhova2019}
N.~Shakhova, I.~Semiletov, and E~Chuvilin.
\newblock Understanding the permafrost–hydrate system and associated methane
  releases in the east siberian arctic shelf.
\newblock {\em Geosciences}, 9:251, 2019.

\bibitem{Bostrom_acsearthspacechem.9b00019}
Mathias Bostr\"om, Robert~W. Corkery, Eduardo R.~A. Lima, Oleksandr~I. Malyi,
  Stefan~Y. Buhmann, Clas Persson, Iver Brevik, Drew~F. Parsons, and Johannes
  Fiedler.
\newblock Dispersion forces stabilize ice coatings at certain gas hydrate
  interfaces that prevent water wetting.
\newblock {\em ACS Earth and Space Chemistry}, 3(6):1014--1022, 2019.

\bibitem{doi:10.1002/2016GL068656}
Nariman Mahabadi, Xianglei Zheng, and Jaewon Jang.
\newblock The effect of hydrate saturation on water retention curves in
  hydrate-bearing sediments.
\newblock {\em Geophysical Research Letters}, 43(9):4279--4287, 2016.

\bibitem{doi:10.1029/2007GC001920}
Tae-Hyuk Kwon, Gye-Chun Cho, and J.~Carlos Santamarina.
\newblock {Gas hydrate dissociation in sediments: Pressure-temperature
  evolution}.
\newblock {\em Geochemistry, Geophysics, Geosystems}, 9(3):Q03019, 2008.

\bibitem{NaturalGas2003}
M.~D. Max, editor.
\newblock {\em {Natural Gas Hydrate in Oceanic and Permafrost Environments}}.
\newblock Kluwer Academic Publishers, Washington DC, USA, 2003.

\bibitem{Matson2018}
D.~L. Matson, A.~G. Davies, T.~V. Johnson, J-P. Coimbe, T.~B. McCord,
  J.~Radebaugh, and S.~Singh.
\newblock {Enceladus' near-surface CO 2 gas pockets and surface frost
  deposits}.
\newblock {\em Icarus}, 302:18--26, 2018.

\bibitem{KamataNimmoSekineKuramotoNoguchiKimuraTani2019}
Shunichi Kamata, Francis Nimmo, Yasuhito Sekine, Kiyoshi Kuramoto, Naoki
  Noguchi, Jun Kimura, and Atsushi Tani.
\newblock {Pluto's ocean is capped and insulated by gas hydrates}.
\newblock {\em Nature Geoscience}, 12(6):407--410, jun 2019.

\bibitem{Sternaad1815}
S.~A. Stern, F.~Bagenal, K.~Ennico, G.~R. Gladstone, W.~M. Grundy, W.~B.
  McKinnon, J.~M. Moore, C.~B. Olkin, J.~R. Spencer, H.~A. Weaver, L.~A. Young,
  T.~Andert, J.~Andrews, M.~Banks, B.~Bauer, J.~Bauman, O.~S. Barnouin,
  P.~Bedini, K.~Beisser, R.~A. Beyer, S.~Bhaskaran, R.~P. Binzel, E.~Birath,
  M.~Bird, D.~J. Bogan, A.~Bowman, V.~J. Bray, M.~Brozovic, C.~Bryan, M.~R.
  Buckley, M.~W. Buie, B.~J. Buratti, S.~S. Bushman, A.~Calloway, B.~Carcich,
  A.~F. Cheng, S.~Conard, C.~A. Conrad, J.~C. Cook, D.~P. Cruikshank, O.~S.
  Custodio, C.~M. Dalle~Ore, C.~Deboy, Z.~J.~B. Dischner, P.~Dumont, A.~M.
  Earle, H.~A. Elliott, J.~Ercol, C.~M. Ernst, T.~Finley, S.~H. Flanigan,
  G.~Fountain, M.~J. Freeze, T.~Greathouse, J.~L. Green, Y.~Guo, M.~Hahn, D.~P.
  Hamilton, S.~A. Hamilton, J.~Hanley, A.~Harch, H.~M. Hart, C.~B. Hersman,
  A.~Hill, M.~E. Hill, D.~P. Hinson, M.~E. Holdridge, M.~Horanyi, A.~D. Howard,
  C.~J.~A. Howett, C.~Jackman, R.~A. Jacobson, D.~E. Jennings, J.~A. Kammer,
  H.~K. Kang, D.~E. Kaufmann, P.~Kollmann, S.~M. Krimigis, D.~Kusnierkiewicz,
  T.~R. Lauer, J.~E. Lee, K.~L. Lindstrom, I.~R. Linscott, C.~M. Lisse, A.~W.
  Lunsford, V.~A. Mallder, N.~Martin, D.~J. McComas, R.~L. McNutt, D.~Mehoke,
  T.~Mehoke, E.~D. Melin, M.~Mutchler, D.~Nelson, F.~Nimmo, J.~I. Nunez,
  A.~Ocampo, W.~M. Owen, M.~Paetzold, B.~Page, A.~H. Parker, J.~W. Parker,
  F.~Pelletier, J.~Peterson, N.~Pinkine, M.~Piquette, S.~B. Porter,
  S.~Protopapa, J.~Redfern, H.~J. Reitsema, D.~C. Reuter, J.~H. Roberts, S.~J.
  Robbins, G.~Rogers, D.~Rose, K.~Runyon, K.~D. Retherford, M.~G.
  Ryschkewitsch, P.~Schenk, E.~Schindhelm, B.~Sepan, M.~R. Showalter, K.~N.
  Singer, M.~Soluri, D.~Stanbridge, A.~J. Steffl, D.~F. Strobel, T.~Stryk,
  M.~E. Summers, J.~R. Szalay, M.~Tapley, A.~Taylor, H.~Taylor, H.~B. Throop,
  C.~C.~C. Tsang, G.~L. Tyler, O.~M. Umurhan, A.~J. Verbiscer, M.~H. Versteeg,
  M.~Vincent, R.~Webbert, S.~Weidner, G.~E. Weigle, O.~L. White,
  K.~Whittenburg, B.~G. Williams, K.~Williams, S.~Williams, W.~W. Woods, A.~M.
  Zangari, and E.~Zirnstein.
\newblock The pluto system: Initial results from its exploration by new
  horizons.
\newblock {\em Science}, 350(6258), 2015.

\bibitem{StuartRossTaylor1999}
S.~R. Taylor.
\newblock {On the difficulties of making Earth-like planets}.
\newblock {\em Meterorites \& Planetary Science}, 34:317--329, 1999.

\bibitem{Elbaum}
Michael Elbaum and M.~Schick.
\newblock {Application of the theory of dispersion forces to the surface
  melting of ice}.
\newblock {\em Phys. Rev. Lett.}, 66:1713--1716, Apr 1991.

\bibitem{doi:10.1029/94GL01858}
Gerald~R. Dickens and Mary~S. Quinby-Hunt.
\newblock Methane hydrate stability in seawater.
\newblock {\em Geophysical Research Letters}, 21(19):2115--2118, 1994.

\bibitem{LUENGO2019}
Juan~Luengo M\'arquez.
\newblock {Fuerzas de van der Waals en la superficie del hielo}.
\newblock Degree Thesis, Universidad Complutense de Madrid, 2019.

\bibitem{JohannesWater2019}
Johannes Fiedler, Mathias Bostr\"om, Clas Persson, Iver~H. Brevik,
  Robert~William Corkery, Stefan~Yoshi Buhmann, and Drew~Francis Parsons.
\newblock Full-spectrum high resolution modeling of the dielectric function of
  water.
\newblock {\em The Journal of Physical Chemistry B}, 124(15):3103--3113, 2020.

\bibitem{Nguyen2015}
Andrew~H. Nguyen, Matthew~A. Koc, Tricia~D. Shepherd, and Valeria Molinero.
\newblock Structure of the ice–clathrate interface.
\newblock {\em The Journal of Physical Chemistry C}, 119(8):4104--4117, 2015.

\bibitem{Ref12}
J.~Daniels.
\newblock {Bestimmung der optischen Konstanten von Eis aus Energie -
  Verlustmessungen von schnellen Elektronen}.
\newblock {\em Optics Communications}, 3(4):240 -- 243, 1971.

\bibitem{Sloan}
E.~Sloan~Jr. and C.~Koh.
\newblock {\em Clathrate Hydrates of Natural Gases}.
\newblock CRC Press, Boca Raton, 2007.

\bibitem{NINHAM201725}
Barry~W. Ninham, Richard~M. Pashley, and Pierandrea~Lo Nostro.
\newblock Surface forces: Changing concepts and complexity with dissolved gas,
  bubbles, salt and heat.
\newblock {\em Current Opinion in Colloid \& Interface Science}, 27:25 -- 32,
  2017.

\bibitem{doi:10.1021/acs.langmuir.0c00117}
Mehdi Azadi, Anh~V. Nguyen, and Gleb~E. Yakubov.
\newblock The effect of dissolved gases on the short-range attractive force
  between hydrophobic surfaces in the absence of nanobubble bridging.
\newblock {\em Langmuir}, 36(34):9987--9992, 2020.

\bibitem{Wettlaufer}
J.~S. Wettlaufer.
\newblock {Impurity Effects in the Premelting of Ice}.
\newblock {\em Physical Review Letter}, 82:2516--2519, Mar 1999.

\bibitem{Prasad}
P.~S.~R. Prasad and B.~S. Kiran.
\newblock Self-preservation and stability of methane hydrates in the presence
  of nacl.
\newblock {\em Scientific Reports}, 9:5860, 2019.

\bibitem{Thiyam2018}
P.~Thiyam, J.~Fiedler, S.~Y. Buhmann, C.~Persson, I.~Brevik, M.~Bostr{\"o}m,
  and D.~F. Parsons.
\newblock {Ice Particles Sink below the Water Surface Due to a Balance of Salt,
  van der Waals, and Buoyancy Forces}.
\newblock {\em The Journal of Physical Chemistry C}, 122(27):15311--15317,
  2018.

\bibitem{Armitage2020}
Thomas W.~K. Armitage, Georgy~E. Manucharyan, Alek~A. Petty, Ron Kwok, and
  Andrew~F. Thompson.
\newblock Enhanced eddy activity in the beaufort gyre in response to sea ice
  loss.
\newblock {\em Nature Communications}, 11:761, 2020.

\bibitem{Hall1988NaClEffects}
D.~L. Hall, S.~M. Sterner, and J.~R. Bodnar.
\newblock Freezing point depression of nacl-kcl-h2o solutions.
\newblock {\em Economic Geology}, 83:197--202, 1988.

\bibitem{Dzya}
I.E. Dzyaloshinskii, E.M. Lifshitz, and L.P. Pitaevskii.
\newblock {The general theory of van der Waals forces}.
\newblock {\em Advances in Physics}, 10(38):165--209, 1961.

\bibitem{DelRio2005}
Frank~W DelRio, Maarten~P de~Boer, James~A Knapp, E~{David Reedy}, Peggy~J
  Clews, and Martin~L Dunn.
\newblock {The role of van der Waals forces in adhesion of micromachined
  surfaces}.
\newblock {\em Nature Materials}, 4(8):629--634, 2005.

\bibitem{Stange2019}
Alexander Stange, Matthias Imboden, Josh Javor, Lawrence~K Barrett, and David~J
  Bishop.
\newblock {Building a Casimir metrology platform with a commercial MEMS
  sensor}.
\newblock {\em Microsystems {\&} Nanoengineering}, 5(1):14, dec 2019.

\bibitem{Autumn2000}
Kellar Autumn, Yiching~A Liang, S~Tonia Hsieh, Wolfgang Zesch, Wai~Pang Chan,
  Thomas~W Kenny, Ronald Fearing, and Robert~J Full.
\newblock {Adhesive force of a single gecko foot-hair}.
\newblock {\em Nature}, 405(6787):681--685, 2000.

\bibitem{Ellingsen_2007}
Simen~A Ellingsen.
\newblock Casimir attraction in multilayered plane parallel magnetodielectric
  systems.
\newblock {\em Journal of Physics A: Mathematical and Theoretical},
  40(9):1951--1961, feb 2007.

\bibitem{Esteso4layerPCCP2020}
V.~Esteso, S.~Carretero-Palacios, L.~G. MacDowell, Johannes Fiedler, D.~F.
  Parsons, F.~Spallek, H.~Míguez, C.~Persson, Stefan~Yoshi Buhmann, I.~Brevik,
  and M.~Bostr\"om.
\newblock Premelting of ice adsorbed on a rock surface.
\newblock {\em Phys. Chem. Chem. Phys.}, 22:11362--11373, 2020.

\bibitem{Buhmann12a}
S.~Y. Buhmann.
\newblock {\em {Dispersion Forces I: Macroscopic quantum electrodynamics and
  ground-state Casimir, Casimir--Polder and van der Waals forces}}.
\newblock Springer, Heidelberg, 2012.

\bibitem{Pars}
V.~A. Parsegian.
\newblock {\em Van der Waals forces: A handbook for biologists, chemists,
  engineers, and physicists}.
\newblock Cambridge University Press, New York, 2006.

\bibitem{Waite2006}
J.~H. Waite, M.~R. Combi, W.~H. Ip, T.~E. Cravens, R.~L. McNutt, W.~Kasprzak,
  R.~Yelle, J.~Luhmann, H.~Niemann, D.~Gell, B.~Magee, G.~Fletcher, G.~Lunine,
  and W.~L Tseng.
\newblock {Cassini ion and neutral mass spectrometer: Enceladus plume
  composition and structure}.
\newblock {\em Science}, 311(5766):1419--1422, 2006.

\bibitem{Postberg2011}
F.~Postberg, J.~Schmidt, Hillier J., S.~Kempf, and R.~Srama.
\newblock {A salt-water reservoir as the source of a compositionally stratified
  plume on Enceladus}.
\newblock {\em Nature}, 474(7353):620--622, 2011.

\bibitem{Bouquet2015}
Alexis. Bouquet, Olivier Mousis, J.~Hunter Waite, and Sylvain Picaud.
\newblock {Possible evidence for a methane source in Enceladus' ocean}.
\newblock {\em Geophysical Research Letters}, 42:1334--1339, 2015.

\bibitem{Matson2012}
D.~L. Matson, J.~C. Castillo-Rogez, A.~G. Davies, and T.~V. Johnson.
\newblock {Enceladus: A hypothesis for bringing both heat and chemicals to the
  surface}.
\newblock {\em Icarus}, 221(1):53--62, 2012.

\bibitem{McKay2003}
C.~P. McKay, K.~P. Hand, P.~T. Doran, D.~T. Andersen, and J.~C. Priscu.
\newblock Clathrate formation and the fate of noble and biologically useful
  gases in lake vostok, antarctica.
\newblock {\em Geophysical Research Letters}, 30(13):35--1--35--4, 2003.

\bibitem{PrietoBallesteros2005}
O.~Prieto-Ballesteros, J.~S. Kargel, M.~Fernandez-Sampedro, F.~Selsis, E.~S.
  Martinez, and D.~L. Hogenboom.
\newblock Evaluation of the possible presence of clathrate hydrates in europas
  icy shell or seafloor.
\newblock {\em Icarus}, 177:491--505, 2005.

\bibitem{Mousis2013}
Olivier Mousis, Azzedine Lakhlifi, Sylvain Picaud, Matthew Pasek, and Eric
  Chassefiere.
\newblock On the abundances of noble and biologically relevant gases in lake
  vostok, antarctica.
\newblock {\em Astrobiology}, 13(4):380--390, 2013.

\bibitem{Safi2017}
E.~Safi, S.~P. Thompson, A.~Evans, S.~J. Day, C.~A. Murray, J.~E. Parker, A.~R.
  Baker, Oliveira~J. M., and J.~Th. van Loon.
\newblock {Properties of CO2 clathrate hydrates formed in the presence of MgSO4
  solutions with implications for icy moons}.
\newblock {\em Astronomy and Astrophysics}, 600(A88):1--9, 2017.

\bibitem{Prachi_concentricice2019}
Prachi Parashar, K.~V. Shajesh, Kimball~A. Milton, Drew~F. Parsons, Iver
  Brevik, and Mathias Bostr\"om.
\newblock Role of zero point energy in promoting ice formation in a spherical
  drop of water.
\newblock {\em Physical Review Research}, 1:033210, Dec 2019.

\bibitem{CRChandbook}
D.~R. Linde, editor.
\newblock {\em {CRC Handbook of Chemistry and Physics (86th ed.)}}.
\newblock CRC Press, Boca Raton (FL), 2005.

\bibitem{Mousis_2009}
Olivier Mousis, Jonathan~I. Lunine, Caroline Thomas, Matthew Pasek, Ulysse
  Marb{\oe}uf, Yann Alibert, Vincent Ballenegger, Daniel Cordier, Yves
  Ellinger, Fran{\c{c}}oise Pauzat, and Sylvain Picaud.
\newblock {CLATHRATION} {OF} {VOLATILES} {IN} {THE} {SOLAR} {NEBULA} {AND}
  {IMPLICATIONS} {FOR} {THE} {ORIGIN} {OF} {TITAN}{\textquotesingle}s
  {ATMOSPHERE}.
\newblock {\em The Astrophysical Journal}, 691(2):1780--1786, feb 2009.

\bibitem{Eistrupetal_AandA2019}
Christian {Eistrup}, Catherine {Walsh}, and Ewine~F. {van Dishoeck}.
\newblock {Cometary compositions compared with protoplanetary disk midplane
  chemical evolution. An emerging chemical evolution taxonomy for comets}.
\newblock {\em Astronomy and Astrophysics}, 629:A84, September 2019.

\bibitem{Postberg2018_book}
F.~Postberg, R.N. Clark, C.J. Hansen, A.J. Coates, C.M. Dalle~Ore, F.~Scipioni,
  M.M. Hedman, and J.H Waite.
\newblock {\em Plume and Surface Composition of Enceladus}, pages 129--162.
\newblock In Schenk et~al. \cite{EnceladusBook}, 2018.

\bibitem{Waite155}
J.~Hunter Waite, Christopher~R. Glein, Rebecca~S. Perryman, Ben~D. Teolis,
  Brian~A. Magee, Greg Miller, Jacob Grimes, Mark~E. Perry, Kelly~E. Miller,
  Alexis Bouquet, Jonathan~I. Lunine, Tim Brockwell, and Scott~J. Bolton.
\newblock Cassini finds molecular hydrogen in the enceladus plume: Evidence for
  hydrothermal processes.
\newblock {\em Science}, 356(6334):155--159, 2017.

\bibitem{LeRoyetal_AandA_2015}
L{\'e}na {Le Roy}, Kathrin {Altwegg}, Hans {Balsiger}, Jean-Jacques
  {Berthelier}, Andre {Bieler}, Christelle {Briois}, Ursina {Calmonte},
  Michael~R. {Combi}, Johan {De Keyser}, Frederik {Dhooghe}, Bj{\"o}rn
  {Fiethe}, Stephen~A. {Fuselier}, S{\'e}bastien {Gasc}, Tamas~I. {Gombosi},
  Myrtha {H{\"a}ssig}, Annette {J{\"a}ckel}, Martin {Rubin}, and Chia-Yu
  {Tzou}.
\newblock {Inventory of the volatiles on comet 67P/Churyumov-Gerasimenko from
  Rosetta/ROSINA}.
\newblock {\em Astronomy and Astrophysics}, 583:A1, November 2015.

\bibitem{JacobsonHujoMolinero_JPCB2009}
Liam~C. Jacobson, Waldemar Hujo, and Valeria Molinero.
\newblock Thermodynamic stability and growth of guest-free clathrate hydrates:
  A low-density crystal phase of water.
\newblock {\em The Journal of Physical Chemistry B}, 113(30):10298--10307,
  2009.

\bibitem{Falenty2014}
Andrzej Falenty, Thomas~C Hansen, and Werner~F Kuhs.
\newblock {Formation and properties of ice XVI obtained by emptying a type sII
  clathrate hydrate}.
\newblock {\em Nature}, 516:231, dec 2014.

\bibitem{Postberg2018}
F.~Postberg, Nozair Khawaja, Bernd Abel, Gael Choblet, Christopher~R. Glein,
  Murthy~S. Gudipati, Bryana~L. Henderson, Hsiang-Wen Hsu, Sascha Kempf, Fabian
  Klenner, Georg Moragas-Klostermeyer, Brian Magee, Lenz N\"olle, Mark Perry,
  Rene Reviol, Jürgen Schmidt, Ralf Srama, Ferdinand Stolz, Gabriel Tobie,
  Mario Trieloff, and J.~Hunter Waite.
\newblock Macromolecular organic compounds from the depths of enceladus.
\newblock {\em Nature}, 558:564, 2018.

\bibitem{Enceladus2018_Ch3_GleinPostbergVance}
C.~R. Glein, F.~Postberg, and S.~D. Vance.
\newblock {The Geochemistry of Enceladus: Composition and Controls}.
\newblock In Paul~M. Schenk, Roger~N Clark, Carly J.~A. Howett, Anne~J.
  Verbiscer, and J.~Hunter Waite, editors, {\em Enceladus and the Icy Moons of
  Saturn}, chapter~3, pages 39--56. The University of Arizona Press, Tucson,
  2018.

\bibitem{POSTBERG2008438}
F.~Postberg, S.~Kempf, J.K. Hillier, R.~Srama, S.F. Green, N.~McBride, and
  E.~Grün.
\newblock The e-ring in the vicinity of enceladus: Ii. probing the moon's
  interior—the composition of e-ring particles.
\newblock {\em Icarus}, 193(2):438 -- 454, 2008.

\bibitem{Postberg2009}
F.~Postberg, S.~Kempf, J.~Schmidt, N.~Brilliantov, A.~Beinsen, B.~Abel,
  U.~Buck, and R.~Srama.
\newblock Sodium salts in e-ring ice grains from an ocean below the surface of
  enceladus.
\newblock {\em Nature}, 459:1098, 2009.

\bibitem{Glein2020}
Christopher~R. Glein and J.~Hunter Waite.
\newblock The carbonate geochemistry of enceladus' ocean.
\newblock {\em Geophysical Research Letters}, 47(3):e2019GL085885, 2020.

\bibitem{Darling1991_NaClEffects}
R.S Darling.
\newblock An extended equation to calculate nacl contents from final clathrate
  melting temperatures in h2o-co2-nacl fluid inclusions: Implications for p-t
  isochore location.
\newblock {\em Geochim. Cosmochim. Acta}, 55:3869–3871, 1991.

\bibitem{Brown1425}
Robert~H. Brown, Roger~N. Clark, Bonnie~J. Buratti, Dale~P. Cruikshank,
  Jason~W. Barnes, Rachel M.~E. Mastrapa, J.~Bauer, S.~Newman, T.~Momary, K.~H.
  Baines, G.~Bellucci, F.~Capaccioni, P.~Cerroni, M.~Combes, A.~Coradini,
  P.~Drossart, V.~Formisano, R.~Jaumann, Y.~Langevin, D.~L. Matson, T.~B.
  McCord, R.~M. Nelson, P.~D. Nicholson, B.~Sicardy, and C.~Sotin.
\newblock Composition and physical properties of enceladus{\textquoteright}
  surface.
\newblock {\em Science}, 311(5766):1425--1428, 2006.

\bibitem{Nicholson509}
Philip~D. Nicholson, Mark~R. Showalter, Luke Dones, Richard~G. French,
  Stephen~M. Larson, Jack~J. Lissauer, Colleen~A. McGhee, Patrick Seitzer,
  Bruno Sicardy, and G.~Edward Danielson.
\newblock Observations of saturn{\textquoteright}s ring-plane crossings in
  august and november 1995.
\newblock {\em Science}, 272(5261):509--515, 1996.

\bibitem{dePater}
I.~de~Pater and J.~J. Lissauer.
\newblock {\em {Planetary sciences, updated 2nd edition}}.
\newblock Cambridge University Press, Cambridge, 2015.

\bibitem{BAUM198184}
W.~A. Baum, T.~Kreidl, J.~A. Westphal, G.~E. Danielson, P.~K. Seidelmann,
  D.~Pascu, and D.~G. Currie.
\newblock Saturn's e ring: I. ccd observations of march 1980.
\newblock {\em Icarus}, 47(1):84 -- 96, 1981.

\bibitem{Hillier2007}
Jon~K. Hillier, S.~F. Green, N.~McBride, J.~P. Schwanethal, F.~Postberg,
  R.~Srama, S.~Kempf, G.~Moragas-Klostermeyer, J.~A.~M. McDonnell, and
  E.~Gr\"un.
\newblock The composition of saturn's e ring.
\newblock {\em Monthly Notices of the Royal Astronomical Society},
  377(4):1588--1596, 2007.

\bibitem{Catling}
D.~C. Catling and J.~F. Kasting.
\newblock {\em {Atmospheric evolution on inhabited and lifeless worlds}}.
\newblock Cambridge University Press, Cambridge, 2017.

\bibitem{Matson2007}
Dennis~L. Matson, Julie~C. Castillo, Jonathan Lunine, and Torrence~V. Johnson.
\newblock Enceladus' plume: Compositional evidence for a hot interior.
\newblock {\em Icarus}, 187(2):569 -- 573, 2007.

\bibitem{Hsu2015}
Hsiang-Wen Hsu, Frank Postberg, Yasuhito Sekine, Takazo Shibuya, Sascha Kempf,
  Mih{\'{a}}ly Hor{\'{a}}nyi, Antal Juh{\'{a}}sz, Nicolas Altobelli, Katsuhiko
  Suzuki, Yuka Masaki, Tatsu Kuwatani, Shogo Tachibana, Sin-iti Sirono, Georg
  Moragas-Klostermeyer, and Ralf Srama.
\newblock {Ongoing hydrothermal activities within Enceladus}.
\newblock {\em Nature}, 519(7542):207--210, mar 2015.

\end{thebibliography}

\end{document}